\documentclass[prb]{revtex4}

\usepackage{graphicx}
\usepackage{dcolumn}
\usepackage{bm}

\begin{document}
\draft
\title{Enhanced transmission versus localization of a light pulse\\
 by a subwavelength metal slit}
\author{S. V.  Kukhlevsky$^a$, M. Mechler$^b$, L. Csap\'o$^c$, K. Janssens$^d$, O. Samek$^e$}
\address{$^a$Institute of Physics, University of P\'ecs, Ifj\'us\'ag u. 6, P\'ecs 7624,
Hungary\\
$^b$South-Trans-Danubian Cooperative Research Centre,
University of P\'ecs, Ifj\'us\'ag u. 6, P\'ecs 7624, Hungary\\
$^c$Institute of Mathematics and Information, University of
P\'ecs,
Ifj\'us\'ag u. 6, P\'ecs 7624, Hungary\\
$^d$Department of Chemistry, University of Antwerp,
Universiteitsplein 1, B-2610 Antwerp, Belgium\\
$^e$Institute of Spectrochemistry and Applied Spectroscopy,
Bunsen-Kirchhoff-Str. 11, D-44139 Dortmund, Germany}
\begin{abstract}
The existence of resonant enhanced transmission and collimation of
light waves by subwavelength slits in metal films [for example,
see T.W. Ebbesen et al., Nature (London) 391, 667 (1998) and H.J.
Lezec et al., Science, 297, 820 (2002)] leads to the basic
question: Can a light pulse be enhanced and simultaneously
localized in space and time by a subwavelength slit? To address
this question, the spatial distribution of the energy flux of an
ultrashort (femtosecond) wave-packet diffracted by a subwavelength
(nanometer-size) slit was analyzed by using the conventional
approach based on the Neerhoff and Mur solution of Maxwell's
equations. The results show that a light pulse can be enhanced by
orders of magnitude and simultaneously localized in the near-field
diffraction zone at the nm- and fs-scales. Possible applications
in nanophotonics are discussed.
\end{abstract}

\maketitle
\section{Introduction}
In the last decade nanostructured optical elements based on
scattering of light waves by subwavelength-size metal objects,
such as particles and screen holes, have been investigated
intensively. The most impressive features of the optical elements
are resonant enhancement and spatial localization of optical
fields by the excitation of electron waves in the metal (for
example, see the
studies~\cite{Neer,Harr,Betz1,Ash,Lewi,Betz2,Pohl,Ebbe,Leze,Port,Hibb,
Gar1,Gar2,Dog,Li,Stoc,Kuk1,Kuk2,Stav,Bori,Taka,Yang,Cschr,Trea,Salo,Cao,Smol,
Barb,Alte,Dykh,Stee,Nawe,Sun,Shi,Scho,Naha,Gome}). Recently some
nanostructures, namely a single subwavelength slit, a grating with
subwavevelength slits, and a subwavelength slit surrounded by
parallel deep and narrow grooves attracted a particular attention
of researchers. The study of resonant enhanced transmission and
collimation of waves in close proximity to a single subwavelength
slit acting as a microscope probe \cite{Neer,Harr,Betz1} was
connected with developing near-field scanning microwave and
optical microscopes with subwavelength resolution
\cite{Ash,Lewi,Betz2,Pohl}. The resonant transmission of light by
a grating with subwavelength slits or a subwavelength slit
surrounded by grooves is an important effect for nanophotonics
\cite{Ebbe,Leze,Port,Hibb}. The transmissivity, on the resonance,
can be orders of magnitude greater than out of the resonance. It
was understood that the enhancement effect has a two-fold origin:
First, the field increases due to a pure geometrical reason, the
coupling of incident plane waves with waveguide mode resonances
located in the slit, and further enhancement arises due to
excitation of coupled surface plasmon polaritons localized on both
surfaces of the slit (grating) \cite{Port,Hibb,Gar1}. A dominant
mechanism responsible for the extraordinary transmission is the
resonant excitation of the waveguide mode in the slit giving a
Fabry-Perot like behaviour \cite{Hibb}. In addition to the
extraordinary transmission, a series of parallel grooves
surrounding a nanometer-size slit can produce a micrometer-size
beam that spreads to an angle of only few degrees \cite{Leze}. The
light collimation, in this case, is achieved by the excitation of
coupled surface plasmon polaritons in the grooves \cite{Gar1}. At
appropriate conditions, a single subwavelength slit flanked by a
finite array of grooves can act as a "lens" focusing light
\cite{Gar2}. It should be noted that the diffractive spreading of
a beam can be reduced also by using a structured aperture or an
effective nanolens formed by self-similar linear chain of metal
nanospheres \cite{Dog,Li}.

New aspects of the problem of resonantly enhanced transmission and
collimation of light are revealed when the nanostructures are
illuminated by an ultra-short (femtosecond) light pulse
\cite{Stoc,Kuk1,Kuk2,Stav,Bori}. For instance, in the study
~\cite{Stoc}, the unique possibility of concentrating the energy
of an ultrafast excitation of an "engineered" or a random
nanosystem in a small part of the whole system by means of phase
modulation of the exciting fs-pulse was predicted. The study
~\cite{Kuk1} theoretically demonstrated the feasibility of
nm-scale localization and distortion-free transmission of fs
visible pulses by a single metal slit, and further suggested the
feasibility of simultaneous super resolution in space and time of
the near-field scanning optical microscopy (NSOM). The
quasi-diffraction-free optics based on transmission of pulses by a
subwavelength nano-slit has been suggested to extend the operation
principle of a 2D NSOM to the "not-too-distant" field regime (up
to $\sim$~0.5 wavelength) \cite{Kuk2}. Some interesting effects,
namely the pulse delay and long living resonant excitations of
electromagnetic fields in the resonant-transition gratings were
recently described in the studies \cite{Stav,Bori}.

The great interest to resonant enhanced transmission, spatial
localization (collimation) of continuous waves and light pulses by
subwavelength metal slits leads to the basic question: Can a light
pulse be enhanced and simultaneously localized in space and time
by a subwavelength slit? If the field enhancement can be achieved
together with nm-scale spatial and fs-scale temporal
localizations, this could greatly increase the potentials of the
nanoslit systems in high-resolution applications, especially in
near-field scanning microscopy and spectroscopy. In the present
article we test whether the resonant enhancement could only be
obtained at the expense of the spatial and temporal broadening of
a light wavepacket. To address this question, the spatial
distribution of the energy flux of an ultrashort (fs) pulse
diffracted by a subwavelength (nanosized) slit in a thick metal
film of perfect conductivity will be analyzed by using the
conventional approach based on the Neerhoff and Mur solution of
Maxwell's equations. In short, we first will summarize the
theoretical development of Neerhoff and Mur (Section II) and the
model will then be used to calculate the spatial distribution of
the energy flux of the transmitted pulse (wavepacket) under
various regimes of the near-field diffraction (Section~III). We
will show that a light pulse can be enhanced by orders of
magnitude and simultaneously localized in the near-field
diffraction zone at the nm- and fs-scales. The implications of the
results for diffraction-unlimited near- and far-field optics will
then be discussed. In Section~IV we summarize the results and
present the conclusions.

\section{Theoretical background}
\begin{figure}
\includegraphics[keepaspectratio,width=13cm]{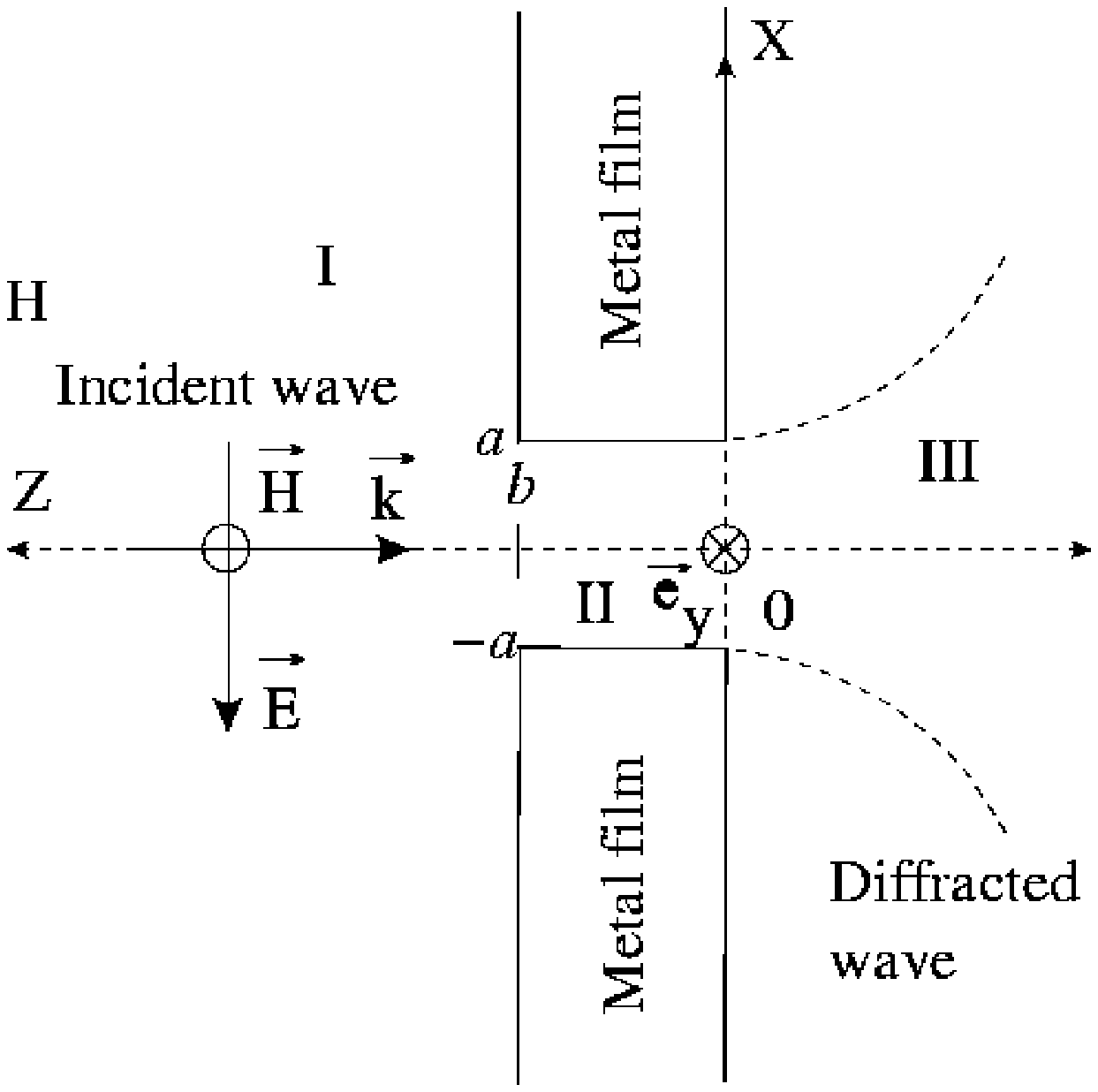}
\caption{\label{vazrajz}Propagation of a continuous wave through a
subwavelength nano-sized slit in a thick metal film.}
\end{figure}

An adequate description of the transmission of light through a
subwavelength nano-sized slit in a thick metal film requires
solution of Maxwell's equations with complicated boundary
conditions.  The light-slit interaction problem even for a
continuous wave can be solved only by extended two-dimensional
$(x,z)$ numerical computations. The three-dimensional $(x,z,t)$
character of the pulse-slit interaction makes the numerical
analysis even more difficult. Let us consider the near-field
diffraction of an ultrashort pulse (wave-packet) by a
subwavelength slit in a thick metal screen of perfect conductivity
by using the conventional approach based on the Neerhoff and Mur
solution of Maxwell's equations. Before presenting a treatment of
the problem for a wave-packet, we briefly describe the Neerhoff
and Mur formulation \cite{Neer,Betz1} for a continuous wave (a
Fourier $\omega$-component of a wavepacket). The transmission of a
plane continuous wave through a slit (waveguide) of width $2a$ in
a screen of thickness $b$ is considered. The perfectly conducting
nonmagnetic screen placed in vacuum is illuminated by a normally
incident plane wave under TM polarization (magnetic-field vector
parallel to the slit), as shown in Fig.~\ref{vazrajz}. The
magnetic field of the wave is assumed to be time harmonic and
constant in the $y$ direction:
\begin{eqnarray}
\label{mfkezd}
{\vec{H}}(x,y,z,t)=U(x,z){\exp}(-i\omega{t}){\vec{e}}_y.
\end{eqnarray}
The electric field of the wave (\ref{mfkezd}) is found from the
scalar field $U(x,z)$ using Maxwell's equations:
\begin{eqnarray}
{E_x}(x,z,t)=-{\frac{ic}{\omega{\epsilon_1}}}{{\partial}_z}{U(x,z)}{\exp}(-i\omega{t}),
\end{eqnarray}
\begin{eqnarray}
{E_y}(x,z,t)=0.
\end{eqnarray}
\begin{eqnarray}
{E_z}(x,z,t)={\frac{ic}{\omega{\epsilon_1}}}{{\partial}_x}{U(x,z)}{\exp}(-i\omega{t}).
\end{eqnarray}
Notice that the restrictions in Eq. \ref{mfkezd} reduce the
diffraction problem to one involving a single scalar field
$U(x,z)$ in two dimensions. The field is represented by
$U_{j}(x,z)$ ($j$=1,2,3 in each of the three regions I, II and
III). The field satisfies the Helmholtz equation:
\begin{eqnarray}
({\nabla}^2+k_{j}^2)U_j=0,
\end{eqnarray}
where $j=1,2,3$. In region I, the field $U_{1}(x,z)$ is decomposed into three
components:
\begin{eqnarray}
\label{U1ossz} U_1(x,z)=U^i(x,z)+U^r(x,z)+U^d(x,z),
\end{eqnarray}
each of which satisfies the Helmholtz equation. $U^i$ represents the incident
field, which is assumed to be a plane wave of unit amplitude:
\begin{eqnarray}
U^i(x,z)=\exp(-ik_1z).
\end{eqnarray}
$U^r$ denotes the field that would be reflected if there were no slit in the screen
and thus satisfies
\begin{eqnarray}
U^r(x,z)=U^i(x,2b-z).
\end{eqnarray}
$U^d$ describes the diffracted field in region I due to the presence of the slit.
With the above set of equations and standard boundary conditions for a perfectly
conducting screen, a unique solution exists for the diffraction problem.
To find the field, the 2-dimensional Green's theorem is applied
with one function given by $U(x,z)$ and the other by a conventional Green's
function:
\begin{eqnarray}
({\nabla}^2+k_{j}^2)G_j=-{\delta}(x-x',z-z'),
\end{eqnarray}
where $(x,z)$ refers to a field point of interest; $x',z'$ are
integration variables, $j=1,2,3$. Since $U_j$ satisfies the
Helmholtz equation, Green's theorem reduces to
\begin{eqnarray}
U(x,z)=\int_{Boundary}(G{\partial}_n{U}-U{\partial}_n{G})dS.
\end{eqnarray}
The unknown fields $U^d(x,z)$, $U_3(x,z)$ and $U_2(x,z)$ are found
using the reduced Green's theorem and boundary conditions on $G$
\begin{eqnarray}
\label{Ud}
U^d(x,z)=-{\frac{\epsilon_1}{\epsilon_2}}\int_{-a}^{a}G_1(x,z;x',b)DU_b(x')dx'
\end{eqnarray}
for $b<z<\infty$,
\begin{eqnarray}
\label{U3}
U_3(x,z)={\frac{\epsilon_3}{\epsilon_2}}\int_{-a}^{a}G_3(x,z;x',0)DU_0(x')dx'
\end{eqnarray}
for $-{\infty}<z<0$,
\begin{eqnarray}
\label{U2} U_2(x,z)=-\int_{-a}^{a}[G_2(x,z;x',0)DU_0(x')-
U_0(x'){\partial}_{z'}G_2(x,z;x',z')|_{z\rightarrow{0^+}}]dx'\nonumber\\
+\int_{-a}^{a}[G_2(x,z;x',b)DU_b(x')-
U_b(x'){\partial}_{z'}G_2(x,z;x',z')|_{z\rightarrow{b^-}}]dx'
\end{eqnarray}
for $|x|<a$ and $0<z<b$. The boundary fields in Eqs.
\ref{Ud}-\ref{U2} are defined by
\begin{eqnarray}
\label{boun1}
U_0(x)=U_2(x,z)|_{z\rightarrow{0^+}},\\
\label{boun2}
DU_0(x)={\partial}_zU_2(x,z)|_{z\rightarrow{0^+}},\\
\label{boun3}
U_b(x)=U_2(x,z)|_{z\rightarrow{b^-}},\\
\label{boun4}
DU_b(x)={\partial}_zU_2(x,z)|_{z\rightarrow{b^-}}.
\end{eqnarray}
In regions I and III the two Green's functions in Eqs. \ref{Ud}
and \ref{U3} are given by
\begin{eqnarray}
G_1(x,z;x',z')={\frac{i}{4}}[H_0^{(1)}(k_1R)+H_0^{(1)}(k_1R')],\\
G_3(x,z;x',z')={\frac{i}{4}}[H_0^{(1)}(k_3R)+H_0^{(1)}(k_3R'')],
\end{eqnarray}
with
\begin{eqnarray}
R=[(x-x')^2+(z-z')^2]^{1/2},\\
R'=[(x-x')^2+(z+z'-2b)^2]^{1/2},\\
R''=[(x-x')^2+(z+z')^2]^{1/2},
\end{eqnarray}
where $H_0^{(1)}$ is the Hankel function. In region II, the
Green's function in Eq. \ref{U2} is given by
\begin{eqnarray}
G_2(x,z;x',z')={\frac{i}{4a{\gamma_0}}}{\exp}(i{\gamma}_0|z-z'|)+
{\frac{i}{2a}}{\sum}_{m=1}^{\infty}{\gamma}_m^{-1}\nonumber\\
{\times}{\cos[m{\pi}(x+a)/2a]}{\cos[m{\pi}(x'+a)/2a]}\nonumber\\
{\times}{\exp(i{\gamma}_m|z-z'|)},
\end{eqnarray}
where $\gamma_m=[k_2^2-(m{\pi}/2a)^2]^{1/2}.$ The field can be
found at any point once the four unknown functions in
Eqs.~\ref{boun1}-\ref{boun4} have been determined. The functions
are completely determined by a set of four integral equations:
\begin{eqnarray}
\label{coup1}
2U_b^i(x)-U_b(x)={\frac{\epsilon_1}{\epsilon_2}}\int_{-a}^{a}G_1(x,b;x',b)DU_b(x')dx',
\end{eqnarray}
\begin{eqnarray}
\label{coup2}
U_0(x)={\frac{\epsilon_3}{\epsilon_2}}\int_{-a}^{a}G_3(x,0;x',0)DU_0(x')dx',
\end{eqnarray}
\begin{eqnarray}
\label{coup3}
{\frac{1}{2}}U_b(x)=-\int_{-a}^{a}[G_2(x,b;x',0)DU_0(x')-
U_0(x'){\partial}_{z'}G_2(x,b;x',z')|_{z\rightarrow{0^+}}]dx'\nonumber\\
+\int_{-a}^{a}[G_2(x,b;x',b)DU_b(x')]dx',
\end{eqnarray}
\begin{eqnarray}
\label{coup4}
{\frac{1}{2}}U_0(x)=\int_{-a}^{a}[G_2(x,0;x',b)DU_b(x')
-U_b(x'){\partial}_{z'}G_2(x,0;x',z')|_{z\rightarrow{b^-}}]dx'\nonumber\\
-\int_{-a}^{a}[G_2(x,0;x',0)DU_0(x')]dx',
\end{eqnarray}
where $|x|<a$, and
\begin{eqnarray}
U_b^i(x)=\exp(-ik_1b).
\end{eqnarray}
The coupled integral equations \ref{coup1}-\ref{coup4}
 for the
four boundary functions are solved numerically. The magnetic
${\vec{H}}(x,z,t)$ and electric ${\vec{E}}(x,z,t)$ fields of the
diffracted wave in region III are found by using Eq.~\ref{U3}. The
fields are given by
\begin{eqnarray}
{\vec{H}}(x,z,t)=i{\frac{a}{N}}{\frac{\epsilon_3}{\epsilon_2}}{\sum_{j=1}^N}
H_0^{(1)}\left[k_{3}\sqrt{(x-x_j)^2+z^2}\right]\nonumber\\
{\times}(D{\vec{U}}_0)_{j}{\exp}(-i\omega{t}){\vec{e}}_y,
\end{eqnarray}
\begin{eqnarray}
E_{x}(x,z,t)=-{\frac{a}{N}}{\frac{\sqrt{\epsilon_3}}{\epsilon_2}}{\sum_{j=1}^N}
{\frac{z}{\sqrt{(x-x_j)^2+z^2}}}H_1^{(1)}\left[k_{3}\sqrt{(x-x_j)^2+z^2}\right]\nonumber\\
{\times}(D{\vec{U}}_0)_{j}{\exp}(-i\omega{t}),
\end{eqnarray}
\begin{eqnarray}
E_{y}(x,z,t)=0,
\end{eqnarray}
\begin{eqnarray}
\label{CWveg}
E_{z}(x,z,t)={\frac{a}{N}}{\frac{\sqrt{\epsilon_3}}{\epsilon_2}}{\sum_{j=1}^N}
{\frac{x-x_j}{\sqrt{(x-x_j)^2+z^2}}}H_1^{(1)}\left[k_{3}\sqrt{(x-x_j)^2+z^2}\right]\nonumber\\
{\times}(D{\vec{U}}_0)_{j}{\exp}(-i\omega{t}),
\end{eqnarray}
where $x_{j}=2a(j-1/2)/N-a$, $j=1,2,...,N$; $N>2a/z$; $H_1^{(1)}$
is the Hankel function. The coefficients $(D{\vec{U}}_0)_{j}$ are
found by solving numerically the four integral equations
\ref{coup1}-\ref{coup4}. For more details of the model and the
numerical solution of the coupled integral equations
\ref{coup1}-\ref{coup4} see refs. \cite{Neer,Betz1}.

Let us now consider the diffraction of an ultra-short pulse (wave
packet). The magnetic field of the incident pulse is assumed to be
Gaussian-shaped in time and both polarized and constant in the $y$
direction:
\begin{eqnarray}
\label{Horig}
{\vec{H}}(x,y,z,t)=U(x,z){\exp[-2\ln(2)(t/{\tau})^2]}{\exp}(-i\omega_{0}{t})
{\vec{e}}_y,
\end{eqnarray}
where $\tau$ is the pulse duration and
$\omega_0=2{\pi}c/\lambda_{0}$ is the central frequency. The pulse
can be composed in the wave-packet form of a Fourier time
expansion (for example, see ref.~\cite{Kuk1,Kuk2}):
\begin{eqnarray}
\label{fourtexp}
{\vec{H}}(x,y,z,t)=\int_{-{\infty}}^{{\infty}}{\vec{H}}(x,z,\omega)
{\exp}(-i\omega{t})d\omega.
\end{eqnarray}
The electric and magnetic fields of the diffracted pulse are found
by using the expressions (\ref{mfkezd}-\ref{CWveg}) for each
Fourier $\omega$-component of the wave-packet (\ref{fourtexp}).
This algorithm is implemented numerically by using the discrete
Fast Fourier-Transform (FFT) instead of the integral
(\ref{fourtexp}). The spectral interval
$[\omega_{min},\omega_{max}]$ and the sampling points $\omega_i$
are optimized by matching the FFT result to the original function
(\ref{Horig}).

The above presented approach deals with the incident waves having
TM polarization. This polarization is considered for the following
reasons. According to the theory of waveguides, the vectorial wave
equations for this polarization can be reduced to one scalar
equation describing the magnetic field H of TM modes. The electric
component E of these modes is found using the field H and
Maxwell's equations. The reduction simplifies the diffraction
problem to one involving a single scalar field in only two
dimensions. The TM scalar equation for the component H is
decoupled from the similar scalar equation describing the field E
of TE (transverse electric) modes. Hence, the formalism works
analogously for TE polarization exchanging the E and H fields.
Moreover, in the case of perfectly conducting nonmagnetic screen
placed in vacuum considered in the present paper, the symmetry of
wave equations indicates that large transmission coefficients
(enhancement effect) do exist at the same experimental conditions
for the TM and TE polarizations.
\section{Numerical analysis and discussion}
\begin{figure}
\includegraphics[keepaspectratio,width=13cm]{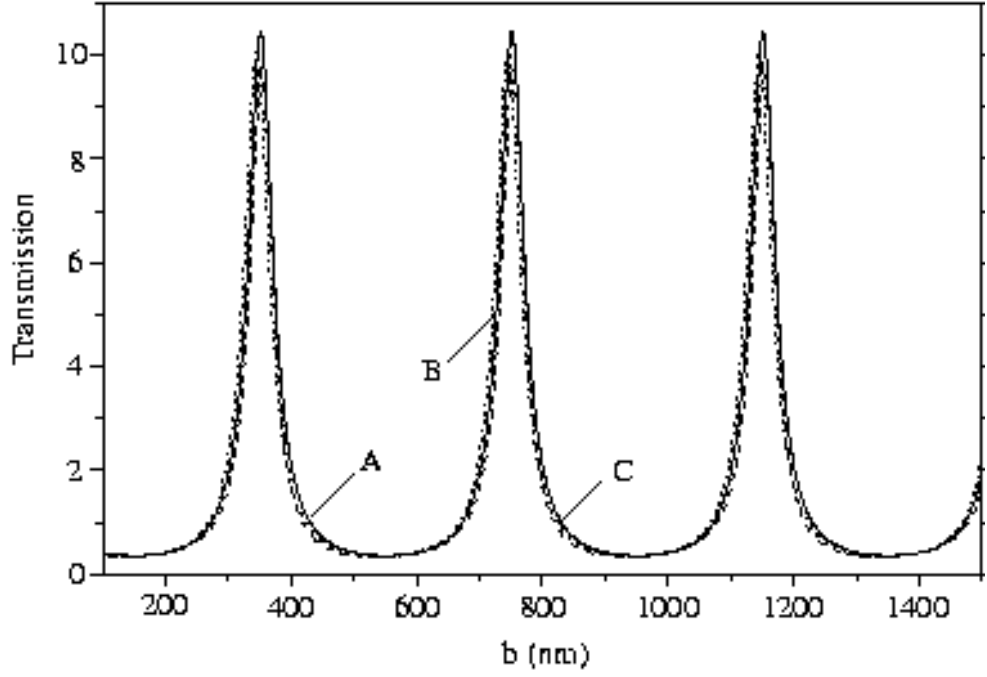}
\caption{\label{Tbdia}The transmission coefficient $T_{cw}$ for a
continuous wave ($\omega$-Fourier component of a wave-packet) as a
function of screen thickness $b$ computed for the wavelength
$\lambda$=800 nm and the slit width $2a$ = 25 nm. Curve A, B and C
correspond to the numerical near-field formula (\ref{nunefi}), the
analytical near-field formula (\ref{annefi}) and the analytical
far-field formula (\ref{anfafi}), respectively. For computational
reasons, in the case of curve C the limits of the integral were
chosen to -1000$\lambda$ and 1000$\lambda$ instead of $-\infty$
and $\infty$; the computation was performed at $z=-10\lambda$. }

\end{figure}

\begin{figure}
\includegraphics[keepaspectratio,width=13cm]{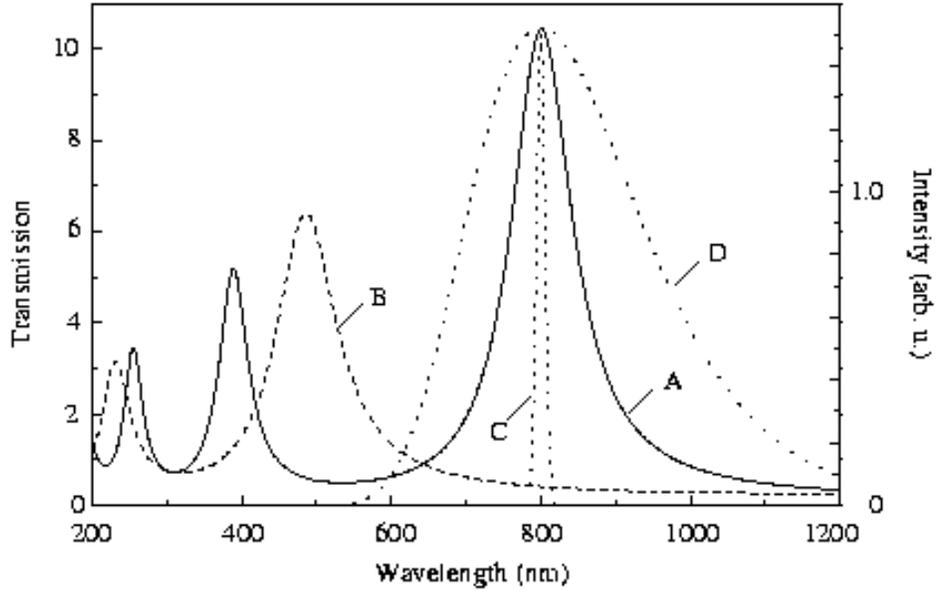}
\caption{\label{Tlamdia}The dispersion $T_{cw}=T_{cw}(\lambda)$
for a continuous wave ($\omega$-Fourier component of a
wave-packet) computed for the slit width $2a$ = 25 nm and
different values of the screen thickness $b$: A~-~350~nm and
B~-~200~nm. The Fourier spectra (curves C and D) are presented for
the comparison. Curves C and D show the Fourier spectra of
incident wave-packets with central wavelength $\lambda_0=$~800~nm
and duration ${\tau}=100$~fs and $\tau=5$~fs, respectively, which
were used in the computations presented in Fig. \ref{ap2tav} and
\ref{atav}.}
\end{figure}

In this section, we test whether a light pulse can be resonantly
enhanced and simultaneously localized in space and time by a
subwavelength nano-sized metal slit. To address this question, the
spatial distribution of the energy flux of the transmitted pulse
under various regimes of the near-field diffraction is analyzed
numerically. The electric ${\vec{E}}$ and magnetic ${\vec{H}}$
fields of the transmitted pulse in the near-field diffraction zone
are computed by solving the equations (\ref{mfkezd}-\ref{CWveg})
for each Fourier $\omega$-component of the wave-packet
(\ref{fourtexp}). The amplitude of a FFT $\omega$-component of the
wave-packet transmitted through the slit depends on the wavelength
$\lambda=2{\pi}c/{\omega}$. Owing to the dispersion, the Fourier
spectra of the transmitted wave-packet changes leading to
modification of the pulse width and duration. The dispersion of a
continuous wave is usually described by the normalized
transmission coefficient $T_{cw}(\lambda)$, which is calculated by
integrating the normalized energy flux $S_z/S_z^i$ over the slit
width~\cite{Neer,Betz1}:
\begin{eqnarray}
\label{nunefi} T_{cw}=-\frac{\sqrt{\epsilon_1}}{4a }\int_{-a}^{a}
{\lim_{z\rightarrow{0^-}}}[(E_{x}H_{y}^*+E_{x}^*H_{y})]dx,
\end{eqnarray}
where $S_z^i$ is the energy flux of the incident wave of unit
amplitude; $S_z$ is the transmitted flux. Equation \ref{nunefi},
which includes the evanescent modes that decay in the far zone,
defines the transmission coefficient in the near-field zone. In
order to establish guidelines for the results of our numerical
analysis, we computed the transmission coefficient
$T_{cw}({\lambda},a,b)$ for a continuous wave (Fourier
$\omega$-component) as a function of screen thickness $b$ and/or
wavelength $\lambda$ for different values of slit width $2a$.
Throughout the computations, the magnitude of the incident
magnetic field is assumed to be equal to 1. We consider a
perfectly conducting nonmagnetic screen placed in vacuum
($\epsilon_1=\epsilon_2=\epsilon_3=1$) As an example, the
dependence $T_{cw}=T_{cw}(b)$ computed for the wavelength
$\lambda$~=~800~nm and the slit width $2a$~=~25~nm is shown in
Fig. \ref{Tbdia}.  The dispersion $T_{cw}=T_{cw}(\lambda)$ for
$2a$~=~25 nm and different values of the screen thickness $b$ is
presented in Fig.~\ref{Tlamdia}. In Fig.~\ref{Tbdia}, we note the
transmission resonances of $\lambda$/2 periodicity with the peak
heights $T_{cw}{\approx}{\lambda}/{2\pi}a$ at the resonances.
Notice that the resonance positions and the peak heights are in
agreement with the results \cite{Harr,Betz1}. The resonance peaks
appear with spacing of 400~nm, but at values somewhat smaller than
400, 800, 1200 nm, etc. In order to better understand the shift of
the resonant wavelengths from the naively expected values we have
derived a simple analytical formula (\ref{annefi}) for the
transmission coefficient $T_{cw}~=~T_{cw}(a,b,\lambda)$ of a
narrow slit (the fundamental mode only is retained):
\begin{eqnarray}
\label{annefi}
T_{cw}(a,b,\lambda)=\frac{a}{2\frac{\pi}{\lambda}}\left[\left(\mathrm{Re}
\left(D\left(b,\lambda\right)\right)\right)^2+
\left(\mathrm{Im}\left(D\left(b,\lambda\right)\right)\right)^2\right]
\end{eqnarray}
, where
\begin{eqnarray}
D(b,\lambda)=\frac{4}{2i\frac{\pi}{\lambda}}\left[\left[
\exp\left(2ib\frac{\pi}{\lambda}\right)\left(A(\lambda)
-\frac{i}{2\pi/\lambda}\right)\right]^2-\left(A(\lambda)+\frac{i}{2\pi/\lambda}\right)^2\right]^{-1}
\end{eqnarray}
and
\begin{eqnarray}
\label{Alam}
A(\lambda)=ia\left[H_0^{(1)}\left(\lambda\right)+\frac{\pi}{2}\left(\mathbf{H}_0\left(\lambda\right)\cdot
H_1^{(1)}\left(\lambda\right)+\mathbf{H}_1\left(\lambda\right)\cdot
H_0^{(1)}]\left(\lambda\right)\right)\right]
\end{eqnarray}
In Eq. \ref{Alam}, $H_0^{(1)}(\lambda)$ and $H_1^{(1)}(\lambda)$
are the Hankel-functions, while $\mathbf{H}_0(\lambda)$ and
$\mathbf{H}_1(\lambda)$ are the Struve-functions. It is
interesting to compare the "near-field" ($z<<\lambda$)
transmittance with the conventional "far-zone" ($z>>\lambda$)
transmittance measured in experiments. The transmittance in the
far-field zone can be described by the following formula:
\begin{eqnarray}
\label{anfafi} T_{cw}=-\frac{\sqrt{\epsilon_1}}{4a
}\int_{-\infty}^{\infty}
{\lim_{z\rightarrow{0^-}}}[(E_{x}H_{y}^*+E_{x}^*H_{y})]dx,
\end{eqnarray}
 Figure~\ref{Tbdia} compares the transmission coefficients $T_{cw} = T_{cw}(b)$ calculated by using the formula (\ref{annefi}) with the values given by the
numerical solution of the equations (\ref{coup1}-\ref{coup4},
\ref{nunefi}) and the values obtained from the evaluation of
(\ref{anfafi}). We notice that the results are practically
undistinguishable. Analysis of the denominator of the formula
(\ref{annefi}) indicates that for $2a/\lambda$ small enough, the
transmission $T_{cw}~=~T_{cw}(b)$ will exhibit the maximums around
wavelengths $\sim$~$\lambda$/2. The shifts of the resonance
wavelengths from the value $\lambda$/2 are caused by the
wavelength dependent terms in the denominator of Eq.~\ref{annefi}.

The dispersion $T_{cw}=T_{cw}(\lambda)$  presented in
Fig.~\ref{Tlamdia} indicates the wave-slit interaction behaviour,
which is similar to those of a Fabry-Perot resonator. The
transmission resonance peaks, however, have a systematic shift
towards longer wavelengths. Analysis of the denominator of the
formula (\ref{annefi}) indicates that for $2a/\lambda$ small
enough, the transmission $T_{cw}~=~T_{cw}(\lambda)$ will exhibit
the Fabry-Perot like maximums around wavelengths where $\sin(kb)$
is zero. Such a behaviour has already been described in refs.
\cite{Hibb,Yang,Taka}. The shifts of the resonance wavelengths
from the Fabry-Perot resonances are caused by the wavelength
dependent terms in the denominator of Eq.~\ref{annefi}. It should
be noted that the values of the shifts calculated using
Eq.~\ref{annefi} are in good agreement with the shifts calculated
using the analytical formula (8) of the study ~\cite{Taka}. Our
computations showed that the peak heights at the main (strongest)
resonant wavelength $\lambda_0^R$ (in the case of
Fig.~\ref{Tlamdia}, $\lambda_0^R$=~500 or 800~nm) are given by
$T_{cw}(\lambda_0^R,a){\approx}{\lambda_0^R}/{2\pi}a$. Notice that
the Fabry-Perot like behaviour of the transmission coefficient is
in agreement with analytical and experimental results published
earlier \cite{Taka,Yang}. The enhancement coefficient ($T_{cw}\sim
30$) calculated using Eqs.~\ref{nunefi} and \ref{annefi}, however,
is different from the attenuation ($T_{cw}~<~1$) predicted by the
study ~\cite{Taka}. The one order difference between the
experimental value and the enhancement calculated in the present
article is attributed to the small transverse dimension $D$
($D\sim\lambda/2$) of the metal plates that form the slit. In our
calculations we studied a slit in infinite transverse dimension
($D~=~\infty$).

Analysis of Fig.~\ref{Tlamdia} indicates that a minimum thickness
of the screen is required to get the waveguide resonance inside
the slit at a given wavelength. The result is in agreement with
the study ~\cite{Taka}, which demonstrated that there is no
transmission peaks at the condition $b~<~\lambda/2$. Notice that
the transmission enhancement mediated by surface plasmons does
exist at considerably smaller thicknesses of the metal in
comparison with the thickness required for the waveguide
resonance. The surface plasmons/polaritons are excited in the
skin, the depth of which is about the extinction length of the
energy decay of electromagnetic wave in the metal. For instance,
the extinction length in an aluminum screen is $\sim$~4~nm for
$\lambda$~=~800~nm. Therefore, for not too narrow slits
(2$a$~=~25~nm) in thick screens ($b$~=~200 and 350~nm) considered
in the paper, the finite skin depth does not fundamentally modify
the process of the extraordinary optical transmission. At the
wavelength 800~nm, the 10-times enhancement
($T_{cw}(a,\lambda)$~$\approx$~$\lambda/2\pi{a}$~$\approx$~10) is
limited by the slit width ($2a$~=~25~nm). In the far-infrared
region ($\lambda$~$\approx$~100~$\mu$m) several orders of
magnitude enhancement can be achieved at the same experimental
conditions. It should be noted that the optimal choice of
parameters has been discussed in the literature and the obtained
enhancement by the factors 10 and 1000 are typical for continuous
waves in the optical and far-infrared spectral ranges. In the case
of narrower and thinner slits, however, the influence of the
finite conductivity effects on the transmission and localization
of a pulse should be taken into account ~\cite{Betz1,Taka}.

\begin{figure}
\includegraphics[keepaspectratio,width=13cm]{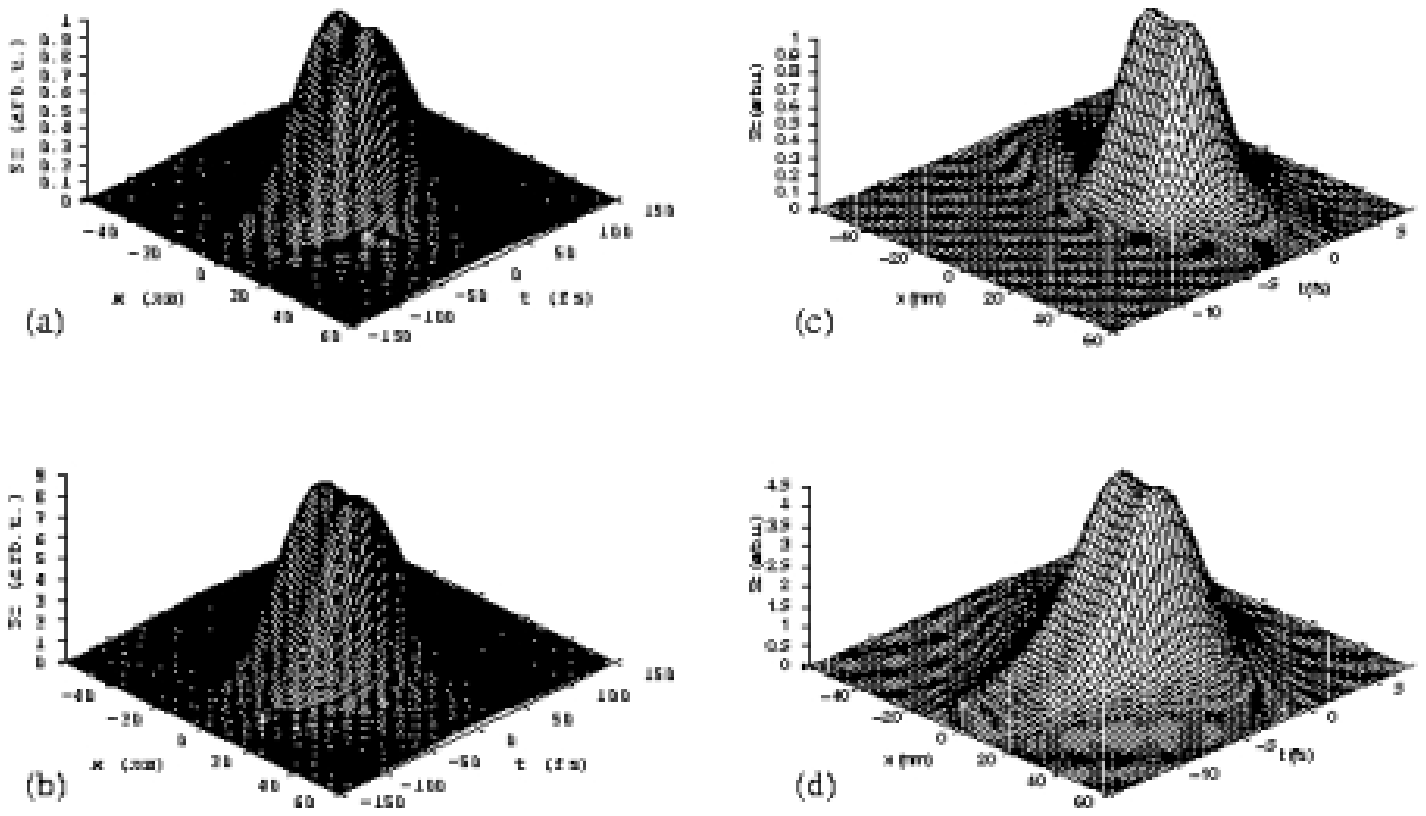}
\caption{\label{ap2tav}The energy flux of a transmitted pulse at
the distance $|z|$ = $a/2$. (a), (c) The non-resonant transmission
by the slit ($2a$~=~25~nm, $b$~=~200~nm; $\tau~=100$~fs and 5~fs,
respectively). (b), (d) The resonant transmission by the slit
($2a$~=~25~nm, $b$~=~350~nm; $\tau=100$~fs and 5~fs,
respectively). The central wavelength of the incident wave-packet
is $\lambda_0=$~800~nm.}
\end{figure}

\begin{figure}
\includegraphics[keepaspectratio,width=13cm]{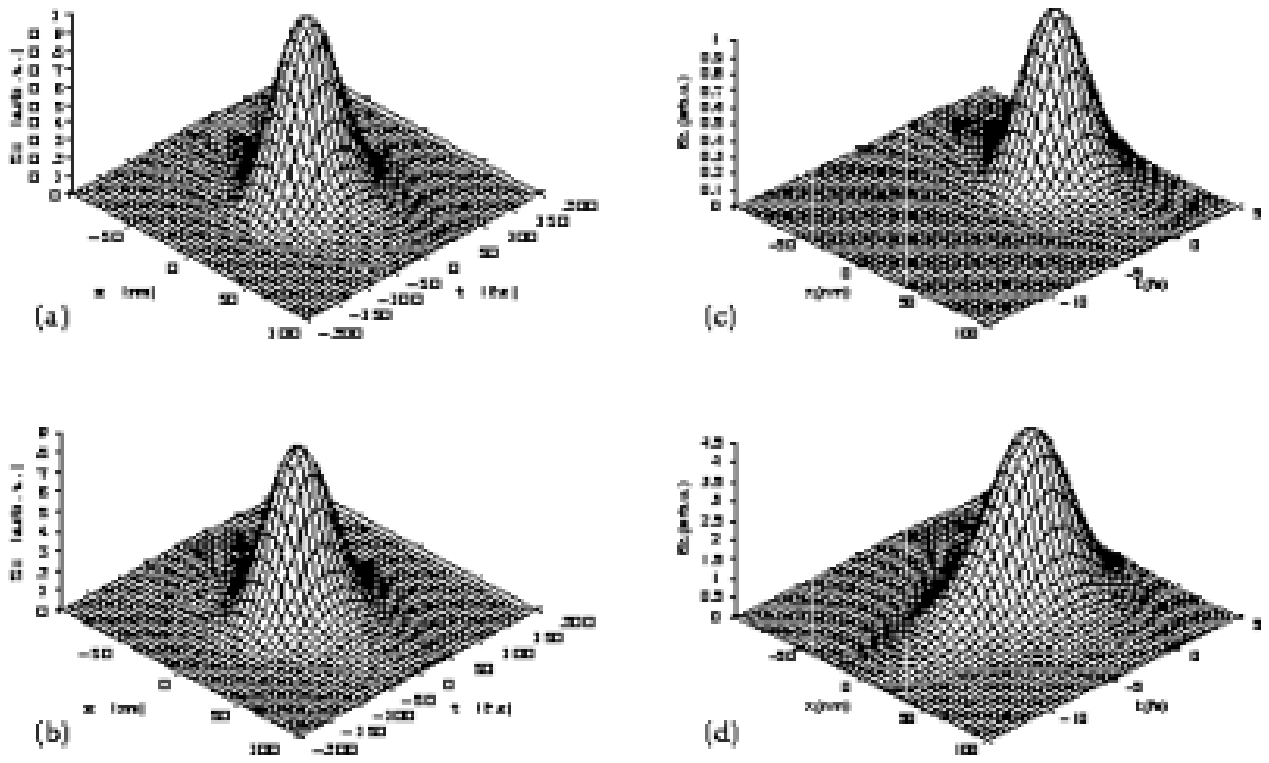}
\caption{\label{atav}The energy flux of a transmitted pulse at the
distance $|z|$~=~$a$. (a), (c) The non-resonant transmission by
the slit ($2a$~=~25 nm, $b$~=~200 nm; $\tau~=100$~fs and 5~fs,
respectively). (b) The resonant transmission by the slit
($2a$~=~25~nm, $b$~=~350~nm; $\tau~=100$~fs and 5~fs,
respectively). The central wavelength of the incident wave-packet
is $\lambda_0=$~800~nm.}
\end{figure}

The existence of transmission resonances for Fourier
$\omega$-components of a wave-packet leads to the question: What
effect the resonant enhancement has on the spatial and temporal
localization of a light pulse? Presumably, the high transmission
at resonance occurs when the system efficiently channels
Fourier-components of the wave-packet from a wide area through the
slit. At resonance, one might assume that if the energy flow is
symmetric about the screen, the pulse width and duration should
increase very rapidly past the screen. Thus the large pulse
strength associated with resonance could only be obtained at the
expense of the spatial and temporal broadening of the wave-packet.
To test this hypothesis, the spatial distributions of the energy
flux of a transmitted wave-packet were computed for different slit
thicknesses corresponding to the resonance and anti-resonance
position. The characteristic difference between the resonant and
non-resonant transmissions could be understood better through a
single figure of the field distributions in all regions I, II and
III. However, it seems to be impossible to do this, because the
visualization of a pulse by a single figure requires a
4-dimensional (U, x, z, t) coordinate system or a great number of
3-dimensional figures at different locations and suitable fixed
times in all the regions. Therefore we present the eight
3-dimensional distributions (\ref{ap2tav}(a)-(d) and
\ref{atav}(a)-(d)) only in the zone of main interest (region III).
As an example, Figs.~\ref{ap2tav}~(a) and \ref{atav}~(a) show the
energy flux of the anti-resonantly transmitted pulses.
Figures~\ref{ap2tav}~(b) and \ref{atav}~(b) correspond to the case
of the waveguide-mode resonance in the slit. Figures~\ref{ap2tav}
and \ref{atav} show the transmitted pulses at the distances
$|z|$~=~$a/2$ and $a$, respectively. The rigorous analysis
\cite{Betz1} showed that the number of modes required for the
accurate computation of the transmittance, near-field
distribution, and other optical characteristics of the system is
given by $N~>~2a/z$, where N is the number of the waveguide modes
and z is the distance from the screen. Therefore at the distances
$|z|$~=~$a/2$ and $a$ the calculations required at least 4 and 2
modes, respectively.

The shape and intensity of an output pulse depends on the slit
parameters and the spectral width of the pulse. For narrow slits,
the spectral width of a 100-fs input pulse is smaller compared to
the spectral width of the resonant transmission (see, Fig.
\ref{Tlamdia}). The comparison of the flux distribution presented
in Fig.~\ref{ap2tav}~(a) with that of Fig.~\ref{ap2tav}~(b) shows
that, for the parameter values adopted, a transmitted wavepacket
is enhanced by one order of magnitude and simultaneously localized
in the 25-nm and 100-fs domains of the near-field diffraction
zone. The shapes of the intensity distributions of the output
pulses are very much the same off and on resonance. The figures
differ only in the order of magnitude of $S_z$. Thus at the
distance $|z|$~=~$a/2$, the slit resonantly enhances the intensity
of the pulse without its spatial and temporal broadening. The
result can be easily understood by considering the dispersion
properties of the slit. For the screen thickness $b$~=~200 nm, the
amplitudes of the Fourier-components of the wave-packet, whose
central wavelength $\lambda_0$ is detuned from the main (at
500~nm) resonance, are practically unchanged in the wavelength
region near 800~nm (see, curve~B in Fig.~\ref{Tlamdia}). This
provides the dispersion- and distortion-free non-resonant
transmission of the wave-packet (Figs.~\ref{ap2tav}(a) and
\ref{atav}(a)). In the case of the thicker screen (b~=~350 nm),
the slit transmission experiences strong mode-coupling regime at
the wavelengths near to 800~nm (see curve A of Fig.~\ref{Tlamdia})
that leads to a profound and uniform enhancement of amplitudes of
all of the Fourier $\omega$-components of the wave-packet (see
curve C in Fig.~\ref{Tlamdia}). Thus, the slit resonantly enhances
by one order of magnitude the intensity of the pulse without its
spatial and temporal broadenings (Figs.~\ref{ap2tav}(b) and
\ref{atav}(b)). Also, notice that at the distance $|z|$~=~$a$
(Figs. \ref{atav}(a) and \ref{atav}(b)), both the resonantly and
anti-resonantly transmitted pulses experience natural spatial
broadening in the transverse direction, while their durations are
practically unchanged. The spectral width of a 5-fs input pulse,
however, is bigger than the spectral width of the resonant
transmission (Fig. \ref{Tlamdia}). In this case the durations of
the resonantly transmitted 5-fs pulses are longer in comparison
with the pulses transmitted off the resonance (Figs.
\ref{ap2tav}(c), (d) and respectively \ref{atav}(c), (d)). We also
notice that the enhancement of the intensity of 5-fs pulses is
approximately two times lower in comparison with the 100-fs
pulses. The temporal broadening of the pulse and the decrease of
the enhancement indicates a natural limitation for the
simultaneous temporal and spatial localization of a pulse together
with its enhancement.

By comparing the data for anti-resonant and resonant transmissions
presented in Figs. \ref{ap2tav} and \ref{atav} one can see that at
the appropriate values of the distance $|z|$ and the wave-packet
spectral width $\Delta{\omega}$ the resonance effect does not
influence the spatial and temporal localization of the
wave-packet. To verify this somewhat unexpected result, the FWHMs
of the transmitted pulse in the transverse and longitudinal
directions were calculated for different values of the slit width
$2a$, central wavelength $\lambda_0$ and pulse duration
${\tau}{\approx}1/\Delta{\omega}_p$ as a function of screen
thickness $b$ at two particular near-field distances $|z|$~=~$a/2$
and $a$ from the screen. It was seen that, at the dispersion-free
resonant transmission condition
$\Delta{\omega}_p~<~\Delta{\omega}_r$, the transmitted pulse
indeed does not experience temporal broadening. Thus the temporal
localization associated with the duration $\tau$ of the incident
pulse remains practically unchanged under the transmission. The
value of $\tau$ is determined by the dispersion-free condition
${\tau}{\approx}1/\Delta{\omega}_p>1/\Delta{\omega}_r$, where
$\Delta{\omega}_r=\Delta{\omega}_r(a)$ practically does not depend
on the screen thickness $b$.  We found that the energy flux of the
transmitted wave-packet can be enhanced by a factor
$T_{cw}(\lambda_0^R,a){\approx}{\lambda_0^R}/{2\pi}a$ by the
appropriate adjusting of the screen thickness $b=b(\lambda_0^R)$,
for an example see Figs.~\ref{Tlamdia}-\ref{atav}. Thus the
wave-packet can be enhanced by a factor ${\lambda_0^R}/{2\pi}a$
and simultaneously localized in the time domain at the
${\tau}={\tau}(a)$ scale. It was also seen that the FWHM of the
transmitted pulse in the transverse direction depends on the
wave-packet central wavelength $\lambda_0$ and the distance $z$
from the slit. Nevertheless, the FWHM of the transmitted pulse can
be always reduced to the value $2a$ by the appropriate decreasing
of the distance $|z|$~=~$|z(a)|$ from the screen ($|z|$~=~$a/2$,
in the case of Fig.~\ref{ap2tav}). Thus high transmission can be
achieved without concurrent loss in the degree of temporal and
spatial localizations of the pulse. In retrospect, this result is
reasonable, since the symmetry of the problem for a time-harmonic
continuous wave (Fourier $\omega$-component of a wave-packet) is
disrupted by the presence of the initial and reflected fields in
addition to the diffracted field on one side (Eq.~\ref{U1ossz}).
As the thickness changes, the fields $U^d$ and $U_3$ change only
in magnitude, but the field $U_1$ changes in distribution as well
since it involves the sum of $U^d$ with unchanging fields $U^i$
and $U^r$. At resonance, the distribution of $U_1$ leads to
channeling of the radiation, but the distribution of $U_3$ remains
unaffected. By the appropriate adjusting of the slit-pulse
parameters a light pulse can be enhanced by orders of magnitude
and simultaneously localized in the near-field diffraction zone at
the nm- and fs-scales.

The limitations of the above analysis must be considered before
the results are used for a particular experimental device. The
resonant enhancement with simultaneous nm-scale spatial and
fs-scale temporal localizations of a light by a subwavelength
metal nano-slit is a consequence of the assumption of the screen
perfect conductivity. The slit can be made of perfectly conductive
(at low temperatures) materials. In the context of current
technology, however, the use of conventional materials like metal
films at a room temperature is more practical. Notice that a metal
can be considered as a perfect conductor in the microwave range.
As a general criterion, the perfect conductivity assumption should
remain valid as far as the slit width and the screen thickness
exceed the extinction length for the Fourier $\omega$-components
of a wavepacket within the metal. The light intensity decays in
the metal screen at the rate of $I_s~=~I_0~{\exp}(-b/{\delta})$,
where ${\delta}$~=~${\delta}(\lambda)$ is the extinction length in
the screen. The aluminum has the largest opacity ($\delta<$~11~nm)
in the spectral region $\lambda~>$~100~nm [20]. The extinction
length increases from 11 to~220~nm with decreasing the wavelength
from 100~to~50~nm. Hence, the perfect conductivity is a very good
approximation in a situation involving a relatively thick
($b~>$~25~nm) aluminum screen and a wave-packet of the duration
${\tau}{\approx}1/\Delta{\omega}_p$ having the Fourier components
in the spectral region $\lambda~>$~100~nm. However, in the case of
thinner screens, shorter pulses and smaller central wavelengths of
wave-packets, the metal films are not completely opaque. This
would reduce a value of spatial localization of a pulse due to
passage of the light through the screen in the region away from
the slit. Moreover, the phase shifts of the Fourier components
along the propagation path caused by the skin effect can modify
the enhancement coefficient and temporal localization properties
of the slit.

The above analysis is directly applicable to the two-dimensional
near-field scanning optical microscopy and spectroscopy. Although
the computations were performed in the case of normal incidence,
the preliminary analysis shows that in the case of oblique
incidence the enhancement effect can be kept without further
spatial and temporal broadening of the pulse. In a conventional 2D
NSOM, a subwavelength ($2a~<~\lambda$) slit illuminated by a
continuous wave is used as a near-field ($|z|<a$) light source
providing the nm-scale resolution in space
\cite{Ash,Lewi,Betz1,Pohl}. The non-resonant transmission of fs
pulses could provide ultrahigh resolution of 2D NSOM
simultaneously in space and time \cite{Kuk1,Kuk2}. The
above-described resonantly enhanced transmission together with nm-
and fs-scale localizations in the space and time of a pulse could
greatly increase the potentials of the 2D near-field scanning
optical microscopy and spectroscopy, especially in high-resolution
applications. However, as a model for a NSOM-tip a hole
(quadratic, rectangular or circular) would be more appropriate
than a slit-type waveguide.  The consequences of the limitation of
region II also in y-direction will result in faster attenuation of
the amplitude of the waveguide modes compared to the slit-type
waveguide modes ~\cite{Betz1}. This should affect the enhancement
and spatial and temporal broadening of the pulse. Moreover, the
walls of the 2D slit have to be parallel over all the thickness to
provide the resonance enhancement effect. Pulled NSOM 3D-tips,
however, usually have conical ends. In the case of the tapered
NSOM 3D waveguides one should also take into account the decrease
of enhancement effect with increasing the waveguide taper. It
should be also noted  that the high transmission
($T_{cw}(\lambda_0^R,a){\approx}{\lambda_0^R}/{2\pi}a$) of a pulse
can be achieved without concurrent loss in the temporal and
spatial localizations of the pulse only at short ($|z|=|z(a)|$)
distances from the slit. The presence of a microscopic sample (a
molecule, for example) placed at the short distance in strong
interaction with NSOM slit modifies the boundary conditions. In
the case of strong slit-sample-pulse interaction, which takes
place at the distance $|z|<< 0.1a$, the response function
accounting for the modification of the quantum mechanical
behaviour of the sample should be taken into consideration. The
potential applications of the effect of the resonantly enhanced
transmission together with nm- and fs-scale localizations of a
pulse are not limited to near-field microscopy and spectroscopy.
Broadly speaking, the effect concerns all physical phenomena and
photonic applications involving a transmission of light by a
single subwavelength nano-slit, a grating with subwavevelength
slits and a subwavelength slit surrounded by parallel grooves (see
the
studies~\cite{Neer,Harr,Betz1,Ash,Lewi,Betz2,Pohl,Ebbe,Leze,Port,Hibb,
Gar1,Gar2,Dog,Li,Stoc,Kuk1,Kuk2,Stav,Bori,Taka,Yang,Cschr,Trea,Salo,Cao,Smol,
Barb,Alte,Dykh,Stee,Nawe,Sun,Shi,Scho,Naha,Gome} and references
therein). For instance, the effect could be used for sensors,
communications, optical switching devices and microscopes.

\section{Conclusion}
In the present article we have considered the question whether a
light pulse can be enhanced and simultaneously localized in space
and time by a subwavelength metal nano-slit. To address this
question, the spatial distributions of the energy flux of an
ultrashort (fs) pulse diffracted by a subwavelength (nanosized)
slit in a thick metal screen of perfect conductivity have been
analyzed by using the conventional approach based on the Neerhoff
and Mur solution of Maxwell's equations. The analysis of the
spatial distributions for various regimes of the near-field
diffraction demonstrated that the energy flux of a wavepacket can
be enhanced by orders of magnitude and simultaneously localized in
the near-field diffraction zone at the nm- and fs-scales. The
extraordinary transmission, together with nm- and fs-scale
localizations of the light pulse, make the nano-slit structures
attractive for many photonic purposes, such as sensors,
communication, optical switching devices and NSOM. We also believe
that the addressing of the above-mentioned basic question gains
insight into the physics of near-field resonant diffraction.
\begin{acknowledgments}
This study was supported by the Fifth Framework of the European
Commission (Financial support from the EC for shared-cost RTD
actions: research and technological development projects,
demonstration projects and combined projects. Contract
NG6RD-CT-2001-00602). The authors thank the Computing Services
Centre, Faculty of Science, University of Pecs, for providing
computational resources.
\end{acknowledgments}
\newpage
\newpage
\newpage
\newpage
\newpage
\newpage
\end{document}